\documentclass[twocolumn,superscriptaddress,preprintnumbers,floatfix]{revtex4}
\usepackage{amsmath,amssymb}
\usepackage{graphicx}
\usepackage{epsfig,color}
\usepackage{bm}
\usepackage{verbatim}
\usepackage{amssymb}
\usepackage{hyperref}
\usepackage{color}
\usepackage{bm}
\usepackage[symbol]{footmisc}
\usepackage{ulem}

\usepackage{xcolor}

\newcommand{\ba}{\begin{eqnarray}}
\newcommand{\ea}{\end{eqnarray}}
\begin{document}
\newcommand\redsout{\bgroup\markoverwith{\textcolor{red}{\rule[0.5ex]{2pt}{0.4pt}}}\ULon}

\title{Masses and decay widths of $\Xi_{c/b}$ and $\Xi^\prime_{c/b}$ baryons}

\author{R. Bijker} 
\affiliation{Instituto de Ciencias Nucleares, Universidad Nacional Aut\'onoma de M\'exico, 04510 Ciudad de M\'exico, M\'exico} 

\author{H. Garc{\'i}a-Tecocoatzi}
\affiliation{The Center for High Energy Physics, Kyungpook National 
University, 80 Daehak-ro, Daegu 41566, Korea}

\author{ A. Giachino} 
\affiliation{INFN, Sezione di Genova, Via Dodecaneso 33, 16146 Genova, Italy}
\affiliation{Institute of Nuclear Physics
Polisch Academy of Sciences Radzikowskiego 152, 31-342 Cracow, Poland}

\author{E. Ortiz-Pacheco}
\affiliation{Instituto de Ciencias Nucleares, Universidad Nacional Aut\'onoma de M\'exico, 04510 Ciudad de M\'exico, M\'exico} 

\author{E. Santopinto}\email[corresponding author: ]{elena.santopinto@ge.infn.it}
\affiliation{INFN, Sezione di Genova, Via Dodecaneso 33, 16146 Genova, Italy}

\begin{abstract}
In this article, we present a complete classification of the negative parity $\Xi'_{c/b}$ and $\Xi_{c/b}$ $P$-wave states: 7 belonging to the $SU(3)$ flavor 
sextet and 7 to the flavor anti-triplet,  the calculation of the $\Xi'_{c/b}$ and $\Xi_{c/b}$ strong partial decay widths  into $^2\Sigma_c \bar{K}$, $^2\Xi_c^{'} \pi$, $^2\Sigma_c \bar{K}$, $^4\Xi_c^{'} \pi$, $\Lambda_c^{'} \bar{K}$, $\Xi_c^{} \pi$ and $\Xi_c^{} \eta$ channels both within the Elementary Emission Model (EEM) and the $^3P_0$ model, and the calculation of the electromagnetic decay widths for $\Xi'_{c/b}$ and $\Xi_{c/b}$ radiative decays.
By means of the equal-spacing mass rule and by the analysis of the strong partial decay widths, we suggest possible assignments for the new LHCb $\Xi_{c}(2923)$, $\Xi_{c}(2939)$, and $\Xi_{c}(2965)$ states, as well as for the $\Xi_{c}$'s previously reported by Belle and BaBar. Our results can be tested by future experiments, at LHCb and Belle, disentangling the remaining missing piece of information, {\it i.e.} the quantum numbers. Finally, a comparison is made between a three-quark and a quark-diquark description of $\Xi_c$ states.  Very recently the LHCb collaboration reported the observation of two  new $\Xi_b$ states, namely $\Xi_b(6327)^{0}$ and $\Xi_b(6333)$, in $\Lambda_b^0 K^{-} \pi^{+}$ channel with a statistical significance larger than nine standard deviations. The experimental masses  and widths of these two states  are consistent with our mass and width  predictions for the doublet of D-wave excitations of the $\Xi_b$ system with  $J^{P}_{\Xi_b(6327)^{0}}=\frac{3}{2}^{+}$ and $J^{P}_{\Xi_b(6333)^{0}}=\frac{5}{2}^{+}$.
\end{abstract}


\maketitle

\section{Introduction}
The observation of three negative parity $\Xi^0_{c}$ charmed baryons by the LHCb collaboration \cite{Aaij:2020yyt} represents an important milestone in our understanding of the quark structure of hadrons. As the hadron mass patterns carry information on the way the quarks interact one another, they provide a means of gaining insight into the fundamental binding mechanism of matter at an elementary level. 
Recent reviews of heavy baryons physics can be found in Refs.~\cite{Roberts:2007ni,Crede:2013sze,Cheng:2015iom,Chen:2016spr,Amhis:2019ckw}. 

The Particle Data Group summary table lists a total of 8 neutral 
single-charm cascade baryons and 7 charged ones \cite{PDG}. 
The angular momentum and parity of these states have not been measured yet. 
The assignment of quantum numbers is based on quark model systematics. 
All ground state single-charm baryons have been identified: the flavor 
anti-triplet with $J^P=1/2^+$ consists of the $\Xi^+_c$, $\Xi^0_c$ and $\Lambda^+_c$ baryons; $\Xi^{\prime +}_c$, $\Xi^{\prime 0}_c$, the three charge states of $\Sigma_c$(2455) and $\Omega^0_c$ form the flavor sextet with $J^P=1/2^+$; the two charge states of $\Xi_c$(2645), the three charge states of $\Sigma_c$(2520) and $\Omega^0_c$(2770) form the flavor sextet with $J^P=3/2^+$. 
Only very recently, the LHCb Collaboration has announced  the observation of three negative parity $\Xi^0_{c}$ states in the $\Lambda^+_c K^-$ channel \cite{Aaij:2020yyt}
\begin{eqnarray}
\Xi^0_c(2923) &:& M = 2923.04 \pm 0.25 \pm 0.20 \pm 0.14~{\rm MeV} \, ,
\nonumber\\                 
&& \Gamma = 7.1 \pm 0.8 \pm 1.8~{\rm MeV} \, , \nonumber\\
\Xi^0_c(2939) &:& M = 2938.55 \pm 0.21 \pm 0.17 \pm 0.14~{\rm MeV} \, , 
\nonumber\\ 
&& \Gamma = 10.2 \pm 0.8 \pm 1.1~{\rm MeV} \, , \nonumber
\end{eqnarray}
\begin{eqnarray}
\Xi^0_c(2965) &:& M = 2964.88 \pm 0.26 \pm 0.14 \pm 0.14~{\rm MeV} \, ,
\nonumber\\                 
&& \Gamma = 14.1 \pm 0.9 \pm 1.3~{\rm MeV} \, .  \nonumber
\nonumber
\end{eqnarray}
As observed by the LHCb collaboration, these three states follow an equal-spacing mass rule of around 126 MeV \cite{Aaij:2020yyt} with respect to the excited negative parity $\Omega^0_c$ states \cite{Aaij:2017nav,Yelton:2017qxg}.
This is similar to the equal spacing mass rule for the $SU_{\rm f}(3)$ ground states in the Gell-Mann Okubo and G\"ursey and Radicati mass formulas \cite{Gell-Mann,Okubo,GRformula}. Therefore, LHCb collaboration suggested that the three new negative parity excited $\Xi^0_c$(2923), $\Xi^0_c$(2939) and $\Xi^0_c$(2965) states be assigned to the same flavor $SU_{\rm f}(3)$ multiplet as the negative parity excited $\Omega^0_c$(3050), $\Omega^0_c$(3065) and $\Omega^0_c$(3090), {\it i.e.} the $SU_{\rm f}(3)$ flavor sextet ($\bf 6$). 
The Belle \cite{Belle1,Belle2} and BaBar \cite{Aubert:2007eb} 
collaborations observed $\Xi_{c}(2930)$ in the same $\Lambda^+_c K^-$ channel, 
maybe it could be considered an unresolved combination of two independent states $\Xi_{c}(2923)$ and $\Xi_{c}(2939)$, while the third LHCb state, the $\Xi_{c}(2965)$ could be either the already observed $\Xi_{c}(2970)$ or maybe a completely new state. In the following we work in the hypothesis that 
the third LHCb state, the $\Xi_{c}(2965)$ is the same state as the already observed $\Xi_{c}(2970)$ and hereafter we denote this state with $\Xi_{c}(2965)$.
Moreover, the single-charm cascade baryons of the anti-triplet are denoted as $\Xi_c$, and those of the sextet are indicated by a prime, $\Xi'_c$.
\\
The new experimental results triggered a large number of theoretical studies 
mainly on the $\Xi^\prime_c$ states of the flavor anti-triplet among others 
using quark-diquark approaches  \cite{Ebert:2007nw,Ebert:2011kk,Yamaguchi:2014era,Yang:2020zjl,Lv:2020qi}, molecular states \cite{HongQiang}, LQCD approach \cite{Oka}, QCD sum rules \cite{Zhang:2008pm,Chen:2015kpa,Yang:2020zjl,Agaev:2020fut,Yang:2020zrh,Wang:2010it}, effective chiral lagrangian approach \cite{Wang:2017kfr,Wang:2020gkn}, 
and the constituent quark model \cite{Chen:2016iyi,Shah2018,ZalakShah,Yoshida}. 

Recently, we introduced an equal-spacing mass formula for heavy baryons which was 
used to study the negative-parity $\Omega^0_c$ baryons and to predict the corresponding negative parity $\Omega_b$ states \cite{Santopinto:2018ljf} 
which subsequently were confirmed by the new experimental data from the LHCb collaboration \cite{omegab}. The masses and decay widths were found to be in agreement within the experimental errors. The aim of this article is to apply the same model to analyze the properties of all ground state and $P$-wave $\Xi'_Q$ (sextet) and $\Xi'_Q$ (anti-triplet) baryons, including the mass spectrum and the decay widths for strong and electromagnetic couplings. 

\section{Harmonic oscillator quark model}

In the quark model, $\Xi_Q$ baryons are described as $usQ$ or $dsQ$ configurations, 
{\it i.e.} a combination of a nonstrange quark which can either be $u$ or $d$, a strange quark $s$, and a heavy quark, $Q=c$ or $b$. The total wave function which is a product of an orbital, spin, flavor and color part has to be antisymmetric. Since physical particles form a color singlet, the color part is antisymmetric, and therefore the orbital-spin-flavor part has to be symmetric under the interchange of the light quarks. The flavor part of the two light quarks can be either symmetric (sextet ${\bf 6}$) or antisymmetric (anti-triplet ${\bf \bar{3}}$), see Fig.~\ref{sextet}. Similarly, the spin part can be either symmetric, $\chi_S$ with $S=3/2$ or $\chi_{\lambda}$ with $S=1/2$, or antisymmetric, $\chi_{\rho}$ with $S=1/2$. The states with maximum spin projection ($M_S=S$) are given by   
\ba
S=\frac{1}{2} &:& \chi_\rho = 
(\uparrow\downarrow\uparrow-\downarrow\uparrow\uparrow)/\sqrt{2} ~, 
\nonumber\\
S=\frac{1}{2} &:& \chi_\lambda = (2\uparrow\uparrow\downarrow
-\uparrow\downarrow\uparrow-\downarrow\uparrow\uparrow)/\sqrt{6} ~,
\nonumber\\
S=\frac{3}{2} &:& \chi_S = \uparrow\uparrow\uparrow ~. 
\ea
The orbital part of the ground state is symmetric, $\psi_0$ with $L^P=0^+_S$, and 
that of the first excited state either symmetric, $\psi_{\lambda}$ with $L^P=1^-_{\lambda}$, or antisymmetric, $\psi_{\rho}$ with $L^P=1^-_{\rho}$. Table~\ref{XiQ} shows the wave functions for the $\Xi_Q$ baryons, where we use the notation $\Xi'_Q$ for the sextet and $\Xi_Q$ for the anti-triplet. The total 
angular momentum $J$ satisfies the usual angular momentum coupling rule 
$|L-S| \leq J \leq L+S$. All states are isospin doublets with isospin $I=1/2$. 
The classification scheme of the negative parity $P$-wave single-charm and single-bottom cascade baryons in a three-quark description shows that there are 
in total 14 such states, 7 belonging to the flavor sextet (5 $\lambda$-mode and 2 $\rho$-mode) and another 7 to the flavor anti-triplet (2 $\lambda$-mode 
and 5 $\rho$-mode). 

\begin{figure}[t]
\centering
\setlength{\unitlength}{0.8pt}
\begin{picture}(225,65)(25,45)
\thicklines
\put( 50,100) {\line( 1,0){100}}
\put( 75, 75) {\line( 1,0){ 50}}
\put( 75, 75) {\line( 1,1){ 25}}
\put(100, 50) {\line( 1,1){ 50}}
\put(125, 75) {\line(-1,1){ 25}}
\put(100, 50) {\line(-1,1){ 50}}
\multiput( 50,100)(50,0){3}{\circle*{8}}
\multiput( 75, 75)(50,0){2}{\circle*{8}}
\put(100, 50){\circle*{8}}
\put( 25, 98){$\Sigma_Q$}
\put( 25, 73){$\Xi'_Q$}
\put( 25, 48){$\Omega_Q$}
\put(200, 75) {\line( 1,0){ 50}}
\put(200, 75) {\line( 1,1){ 25}}
\put(250, 75) {\line(-1,1){ 25}}
\put(225,100) {\circle*{8}}
\multiput(200, 75)(50,0){2}{\circle*{8}}
\put(175, 98){$\Lambda_Q$}
\put(175, 73){$\Xi_Q$}
\end{picture}
\caption{Heavy baryon sextet (left) and anti-triplet(right).}
\label{sextet}
\end{figure}
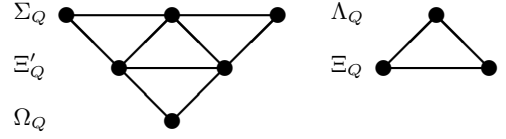

\begin{table}[t]
\centering
\caption{Classification of highest charge state of sextet $\Xi'_Q$ baryons (top) 
and anti-triplet $\Xi_Q$ baryons (bottom). The upper index in the first column denotes the spin degeneracy, $2S+1$.}
\label{XiQ}
\vspace{10pt}
\begin{tabular}{ccccc}
\hline\hline
\noalign{\smallskip}
State & Wave function & $(n_{\rho},n_{\lambda})$ & $L^P$ & $J^P$ \\
\noalign{\smallskip}
\hline
\noalign{\smallskip}
$^{2}\Xi'_Q$ & $\frac{1}{\sqrt{2}}(us+su)Q 
\left[ \psi_{0} \chi_{\lambda} \right]$ 
& $(0,0)$ & $0^+$ & $\frac{1}{2}^+$ \\ 
\noalign{\smallskip}
$^{4}\Xi'_Q$ & $\frac{1}{\sqrt{2}}(us+su)Q 
\left[ \psi_{0} \chi_{S} \right]$ 
& $(0,0)$ & $0^+$ & $\frac{3}{2}^+$ \\ 
\noalign{\smallskip}
$^{2}\lambda(\Xi'_Q)_J$ & $\frac{1}{\sqrt{2}}(us+su)Q 
\left[ \psi_{\lambda} \chi_{\lambda} \right]_J$ 
& $(0,1)$ & $1^-$ & $\frac{1}{2}^-$, $\frac{3}{2}^-$ \\ 
\noalign{\smallskip}
$^{4}\lambda(\Xi'_Q)_J$ & $\frac{1}{\sqrt{2}}(us+su)Q 
\left[ \psi_{\lambda} \chi_{S} \right]_J$  
& $(0,1)$ & $1^-$ & $\frac{1}{2}^-$, $\frac{3}{2}^-$, $\frac{5}{2}^-$ \\ 
\noalign{\smallskip}
$^{2}\rho(\Xi'_Q)_J$ & $\frac{1}{\sqrt{2}}(us+su)Q 
\left[ \psi_{\rho} \chi_{\rho} \right]_J$  
& $(1,0)$ & $1^-$ & $\frac{1}{2}^-$, $\frac{3}{2}^-$ \\ 
\noalign{\smallskip}
\hline
\noalign{\smallskip}
$^{2}\Xi_Q$ & $\frac{1}{\sqrt{2}}(us-su)Q 
\left[ \psi_{0} \chi_{\rho} \right]$ 
& $(0,0)$ & $0^+$ & $\frac{1}{2}^+$ \\ 
\noalign{\smallskip}
$^{2}\lambda(\Xi_Q)_J$ & $\frac{1}{\sqrt{2}}(us-su)Q 
\left[ \psi_{\lambda} \chi_{\rho} \right]_J$ 
& $(0,1)$ & $1^-$ & $\frac{1}{2}^-$, $\frac{3}{2}^-$ \\ 
\noalign{\smallskip}
$^{2}\rho(\Xi_Q)_J$ & $\frac{1}{\sqrt{2}}(us-su)Q 
\left[ \psi_{\rho} \chi_{\lambda} \right]_J$  
& $(1,0)$ & $1^-$ & $\frac{1}{2}^-$, $\frac{3}{2}^-$ \\ 
\noalign{\smallskip}
$^{4}\rho(\Xi_Q)_J$ & $\frac{1}{\sqrt{2}}(us-su)Q 
\left[ \psi_{\rho} \chi_{S} \right]_J$  
& $(1,0)$ & $1^-$ & $\frac{1}{2}^-$, $\frac{3}{2}^-$, $\frac{5}{2}^-$ \\ 
\noalign{\smallskip}
\hline\hline
\end{tabular}
\end{table}

Following Ref.~\cite{Santopinto:2018ljf} we consider a harmonic oscillator quark model with a spin, spin-orbit, isospin and flavor dependent terms  
\ba
M = H_{\rm ho} + A \, \vec{S} \cdot \vec{S} + B \, \vec{L} \cdot \vec{S} 
+ E \, \vec{I} \cdot \vec{I} + G \, C_{2SU_{\rm f}(3)} ~.
\label{hosc}
\ea
The harmonic oscillator quark model for $qqQ$ baryons with two light quarks 
and one heavy quark is given by \cite{Isgur:1978xj}
\ba
H_{\rm ho} &=& \sum_i \left( m_i + \frac{p_i^2}{2m_i} \right) 
+ \frac{1}{2} C \sum_{i<j} (\vec{r}_i - \vec{r}_j)^2 
\nonumber\\ 
&=& M + \frac{P^2}{2M} + \frac{p_{\rho}^2}{2m_{\rho}}  
+ \frac{1}{2} m_{\rho} \omega_{\rho}^2 \rho^2 
\nonumber\\
&& + \frac{p_{\lambda}^2}{2m_{\lambda}} 
+ \frac{1}{2} m_{\lambda} \omega_{\lambda}^2 \lambda^2 ~,
\ea
where we have made a change of variables to relative Jacobi coordinates 
\ba
\vec{\rho} &=& (\vec{r}_1 - \vec{r}_2)/\sqrt{2} ~,
\nonumber\\
\vec{\lambda} &=& (\vec{r}_1 + \vec{r}_2 - 2\vec{r}_3)/\sqrt{6} ~,
\ea
and the center-of-mass coordinate, and their canonically conjugate momenta. 
The labels $1$ and $2$ refer to the light quarks and $3$ to the heavy quark. 

\begin{figure*}[t]
\centering
\begin{minipage}{\linewidth}
\includegraphics[width=0.47\linewidth]{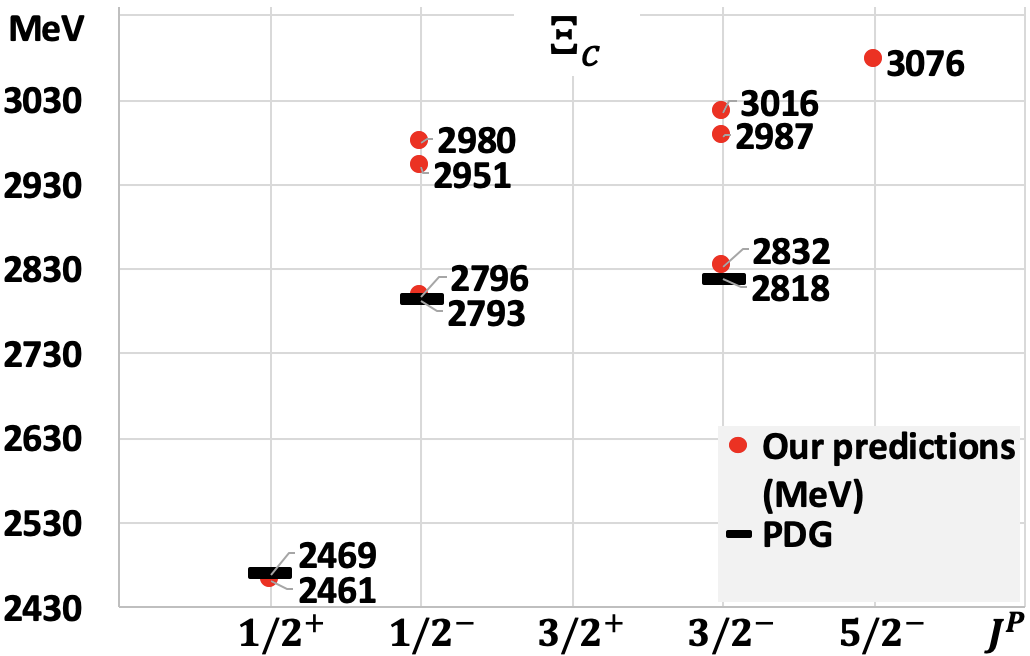}
\hfill
\includegraphics[width=0.48\linewidth]{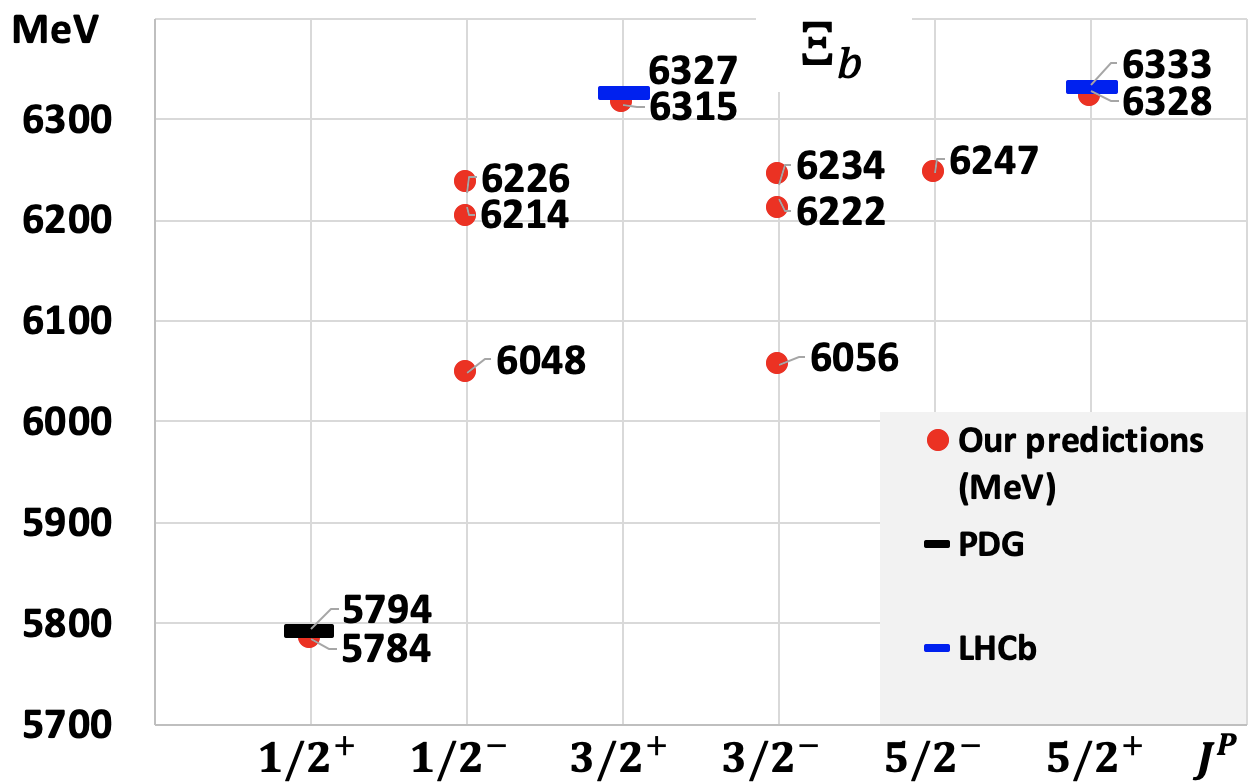}
\end{minipage}
\vfill
\begin{minipage}{\linewidth}
\includegraphics[width=0.47\linewidth]{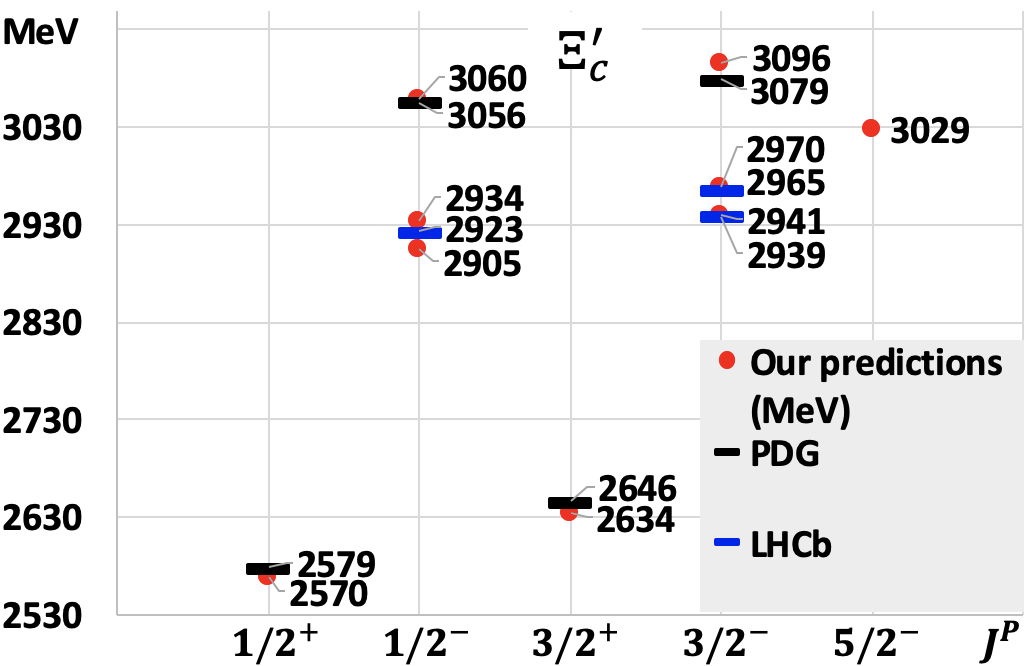}
\hfill
\includegraphics[width=0.47\linewidth]{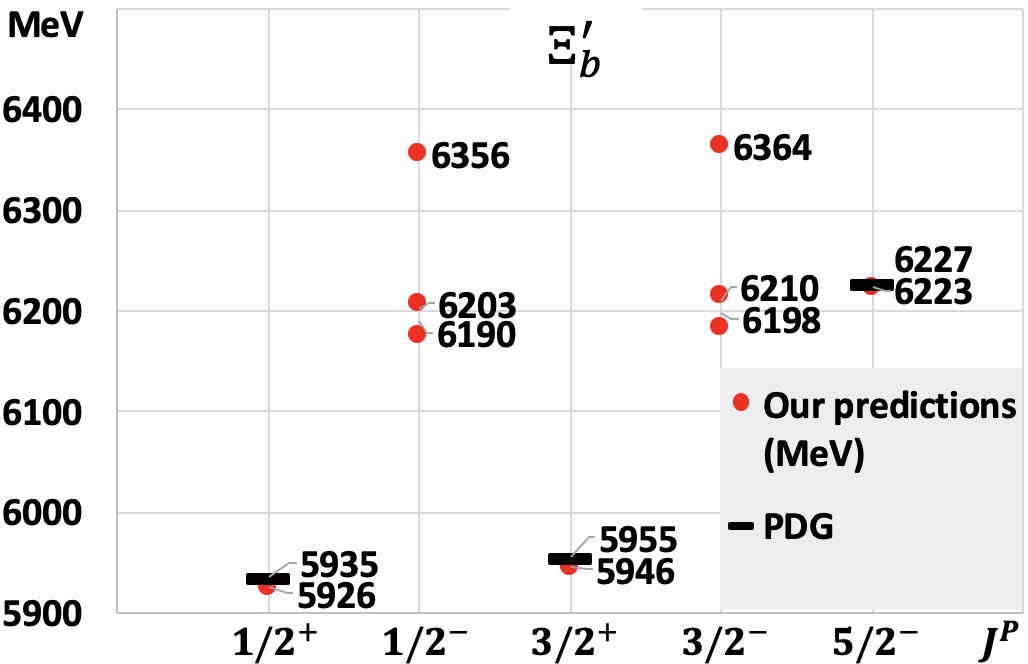}
\end{minipage}
\caption{Mass spectra and tentative quantum number assignments for $\Xi_c$ (top left), $\Xi'_c$ (bottom left), $\Xi_b$ (top right) and $\Xi'_b$ (bottom right). The theoretical  predictions (red dots) are compared with the experimental results from the LHCb collaboration \cite{Aaij:2020yyt} \cite{LHCb:2021ssn} (blue lines) and the Particle Data Group compilation (black lines) \cite{PDG}.}
\label{spectrumXiQ}
\end{figure*}

For $\Xi_Q$ and $\Xi'_Q$ baryons the reduced masses are given by $m_{\rho}=(m_{u/d}+m_s)/2$, {\it i.e} the average of the strange and nonstrange quark masses, and $m_{\lambda}=3 m_{\rho}m_Q/M$ with $M=2m_{\rho}+m_Q$. 
Whereas in the light-baryon phenomenology, the $\rho$- and $\lambda$-excitations are degenerate in energy, in the heavy-light sector they are decoupled with 
frequencies, $\omega_{\rho}=\sqrt{3C/m_{\rho}}$ and $\omega_{\lambda}=\sqrt{3C/m_{\lambda}}$. For $qqQ$ baryons with two light and 
one heavy quark the frequencies satisfy $\omega_{\lambda} < \omega_{\rho}$, whereas for $QQq$ baryons with two heavy and one light quarks the situation is reversed, $\omega_{\rho} < \omega_{\lambda}$. For equal masses, as is the case for $qqq$ and $QQQ$ baryons, the two frequencies become the same, $\omega_{\rho} = \omega_{\lambda}$. \indent
Since all $\Xi'_Q$ and $\Xi_Q$ baryons are isospin doublets with $I=1/2$  
and have the same quark content, the isospin term and the mass term, $M=m_{u/d}+m_s+m_Q$, give the same contribution to all states. As a consequence,  
mass differences only depend on the harmonic oscillator frequencies, $\hbar \omega_{\rho}$ and $\hbar \omega_{\lambda}$, the spin term $A$, the spin-orbit term $B$ and the flavor-dependent term $G$
\ba
M &=& M( ^{2}\Xi_Q) + \hbar \omega_{\rho} n_{\rho} 
+ \hbar \omega_{\lambda} n_{\lambda} + A \left[ S(S+1) - \frac{3}{4} \right]
\nonumber\\
&& + B \frac{1}{2} \left[ J(J+1) - L(L+1) - S(S+1) \right] 
\nonumber\\
&& + G \frac{1}{3} \left[ p(p+3)+q(q+3)+pq-4 \right] ~.
\label{MassFormula}
\ea 
The spin-dependent term splits the states with different spin content, such as the $^2\Xi'_Q$ and $^4\Xi'_Q$ configurations in Table~\ref{XiQ}. The spin-orbit interaction which is small in light baryons \cite{Ebert:2007nw,Capstick:1986bm}, turns out to be fundamental to describe the heavy-light baryon mass patterns \cite{Santopinto:2018ljf}. The effect of the spin-orbit term is to split the states with different $J$ in configurations like $^{2}\lambda(\Xi'_Q)_J$. 
Finally, the flavor-dependent term splits the $\Xi'_Q$ baryons belonging to the flavor sextet, ${\bf 6}$ with $(p,q)=(2,0)$, from the $\Xi_Q$ baryons of the anti-triplet, ${\bf \bar{3}}$ with $(p,q)=(0,1)$. 

There are two types of radial excitations, the $\rho$-mode corresponds to an excitation in the relative coordinate between the two light quarks and the $\lambda$-mode to an excitation in the relative coordinate between the two light quarks and the heavy quark. In quark-diquark models, baryons are described as an effective two-body quark-diquark systems in which the excitation of the $\rho$-mode between the two light quarks is not taken into account. As a consequence, the number of excited states is much smaller than in the three-quark picture. Specifically, Table~\ref{XiQ} shows that in a three-quark model there are 7 excited states for sextet $\Xi'_Q$ baryons, and another 7 for anti-triplet $\Xi_Q$ baryons, whereas in quark-diquark models the $\rho$-mode excitations are absent, leading to 5 excited $\Xi'_Q$ baryons and 2 excited $\Xi_Q$ baryons. Therefore, the identification of $\rho$-mode excitations in the spectrum of heavy-light baryons provides a tool to discriminate between three-quark and quark-diquark descriptions, since the $\rho$-mode excitations are absent in quark-diquark models, but are allowed in three-quark models of baryons \cite{Santopinto:2018ljf}. 

\section{Mass spectrum}
 In the present article we consider single-charm and single-bottom $\Xi_Q$ baryons associated with the ground-state configuration $\psi_0$ with $(n_{\rho},n_{\lambda})=(0,0)$, and with one quantum of excitation, either in the $\lambda$ mode, $\psi_{\lambda}$, with $(n_{\rho},n_{\lambda})=(0,1)$, or in the $\rho$ mode, $\psi_{\rho}$, with $(n_{\rho},n_{\lambda})=(1,0)$. 
The allowed configurations are given in Table~\ref{XiQ}. 

The mass formula of Eqs.~(\ref{hosc}) and (\ref{MassFormula}) is an extension 
of the Gell-Mann-Okubo and G\"ursey-Radicati mass formulas. It was introduced in Ref.~\cite{Santopinto:2018ljf} to study the recently observed negative parity $\Omega_c$ states by the Belle and LHCb collaborations  \cite{Aaij:2017nav,Yelton:2017qxg}. It was shown that these $\Omega_c$ states 
can be interpreted as $\lambda$-mode excitations. In addition, it was used to predict the masses of the corresponding negative parity $\Omega_b$ baryons which subsequently were measured by the LHCb collaboration \cite{omegab}. Here we apply the same mass formula to the $\Xi_Q$ and $\Xi'_Q$ baryons, using the same parameter values as determined in the previous study of $\Omega_Q$ baryons \cite{Santopinto:2018ljf} without any additional fine-tuning or the introduction of extra parameters. 

In this way, the mass spectra of the $\Xi_Q$ and $\Xi'_Q$ baryons are calculated in 
a parameter-free procedure. The results for ground state and $\lambda$- and $\rho$-mode excitations are reported in Fig.~\ref{spectrumXiQ} and in  Tables~\ref{Xictable} and \ref{Xibtable}. 
The assignments are mostly based on systematics of mass spectra. 
For the singly-charm baryons, there is no doubt about the identification of the $^2\Xi'_c$, $^4\Xi'_c$ and $^2\Xi_c$ configurations. The masses of the recently measured $\Xi_c^0$(2923), $\Xi_c^0$(2939) and $\Xi_c^0$(2965) resonances \cite{Aaij:2020yyt}, together with the observed equal spacing rule \cite{Gell-Mann,Okubo,Aaij:2020yyt} with excited $\Omega_c^0$ baryons
\ba
&& M(\Omega_c(3050))-M(\Xi_c(2923) 
\nonumber\\
&& \qquad \simeq M(\Omega_c(3065))-M(\Xi_c(2939) 
\nonumber\\
&& \qquad \simeq M(\Omega_c(3090))-M(\Xi_c(2965) 
\nonumber\\
&& \qquad \simeq 125 \mbox{ MeV} ~,
\ea
suggest the same assignment as for the $\Omega_c$ baryons: $\lambda$-mode 
excitations of flavor sextet configurations $^4\lambda(\Xi'_c)_{1/2^-}$, $^2\lambda(\Xi'_c)_{3/2^-}$ and $^4\lambda(\Xi'_c)_{3/2^-}$, respectively.  
The $\Xi_c$(3055) and $\Xi_c$(3080) resonances are assigned as $\rho$-mode 
excitations of the sextet configuration, $^2\rho(\Xi'_c)_{J^P}$ with $J^P=1/2^-$ 
and $3/2^-$ which would indicate a preference of a three-quark over a quark-diquark description. Finally, the $\Xi_c$(2790) and $\Xi_c$(2818) 
resonances are assigned as $\lambda$-mode excitations of the flavor anti-triplet.  
 
For the singly-bottom $\Xi_b$ baryons the experimental information is more 
scarce. The $^2\Xi'_b$, $^4\Xi'_b$ and $^2\Xi_b$ configurations can be assigned 
without problem. The newly observed $\Xi_b$(6227)$^-$ and $\Xi_b$(6227)$^0$ resonances \cite{Aaij:2018yqz,xibzero} can be 
assigned either as a $\lambda$-mode excitation of the flavor sextet,  $^4\lambda(\Xi'_b)_{5/2^-}$ (as in Fig.~\ref{spectrumXiQ} and Table~\ref{Xibtable}), 
or as a $\rho$-mode excitation of the flavor anti-triplet, $^2\rho(\Xi_b)_{3/2^-}$ 
or $^4\rho(\Xi_b)_{3/2^-}$. Very recently the LHCb collaboration reported the observation of two  new $\Xi_b$ states  in $\Lambda_b^0 K^{-} \pi^{+}$ channel with a statistical significance larger than nine standard deviations \cite{LHCb:2021ssn}.
Our model  of Eq. \ref{hosc}  predicts  the masses of these two states as D-wave excitations of the $\Xi_b$ system with $L=l_{\lambda}=2$ and total angular momentum  $J^{P}_{\Xi_b(6327)^{0}}=\frac{3}{2}^{+}$ and $J^{P}_{\Xi_b(6333)^{0}}=\frac{5}{2}^{+}$ (see the right  upper side of Fig. \ref{spectrumXiQ}).
The predicted masses are respectively 6315 MeV and 6328 MeV for $\Xi_b(6327)^{0}$ and $\Xi_b(6333)^{0}$ in 
 good agreement with the experimental masses measured by the LHCb collaboration \cite{LHCb:2021ssn}: $ m\left(\Xi_{b}(6327)^{0}\right) = 6327.28_{-0.21}^{+0.23} \pm 0.08 \pm 0.24$  MeV and $m\left(\Xi_{b}(6333)^{0}\right) = 6332.69_{-0.18}^{+0.17} \pm 0.03 \pm 0.22$ MeV.

\section{Decay widths}

For the evaluation of strong and electromagnetic decay widths we first discuss the radial wave functions. The harmonic oscillator wave functions depend on the two size parameters, one for the $\rho$ coordinate and one for the $\lambda$ coordinate. The relative wave function for the ground state with $(n_{\rho},n_{\lambda})=(0,0)$ is given by 
\ba
\psi^{\rm o}(\vec{\rho},\vec{\lambda}) = 
\left( \frac{\alpha_\rho \alpha_\lambda}{\pi} \right)^{3/2}   \mbox{e}^{-(\alpha_\rho^2 \vec{\rho}^{\, 2} 
+ \alpha_\lambda^2 \vec{\lambda}^{\, 2})/2} ~,
\ea
with 
\ba
\alpha_\lambda = \alpha_\rho \left( \frac{3m'}{2m+m'} \right)^{1/4} ~.
\label{alphas}
\ea
The oscillator parameter $\alpha_\rho$ used in the calculation of the strong decays within the elementary emission model (EEM) can be determined from the proton charge radius
\ba
\left< r^2_p \right>^{1/2}_{\rm ch} = \frac{1}{\alpha_\rho} ~,
\ea
to give $\alpha_\rho=237$ MeV. The value of $\alpha_\lambda$ can be determined 
from Eq.~(\ref{alphas}) as $\alpha_\lambda=283$ MeV for $\Xi_c$ and $\Xi'_c$ baryons 
and $\alpha_\lambda=301$ MeV for $\Xi_b$ and $\Xi'_b$ baryons. 
In the case of the  $^{3}P_0$ model, we also prefer not to take $\alpha_\rho$ and $\alpha_\lambda$ as free parameters, but to express them in terms of the baryon $\rho$- and $\lambda$-mode frequencies  \cite{Santopinto:2018ljf}, $\omega_{\rho,\lambda}=\sqrt{3K_Q/m_{\rho,\lambda}}$,  using the relation  $\alpha^2_{\rho,\lambda}=\omega_{\rho,\lambda}m_{\rho,\lambda}$  for both initial- and final-state baryon resonances. 

\subsection{Strong couplings}

Strong couplings provide an important test of baryon wave functions, 
and can be used to distinguish between different models of baryon 
structure. Here we consider strong decays of baryons by the emission 
of a pseudoscalar meson
\ba
B \rightarrow B^{\prime} + M ~.
\ea
Several forms have been suggested for the form of the operator inducing 
the strong transition \cite{LeYaouanc}. In this article we consider the 
strong decays of $\Xi_Q$ and $\Xi'_Q$ baryons by the emission of
a pseudoscalar meson as calculated in the elementary emission model (EEM), 
as well as in the $^3P_0$ model. 

\begin{table*}
\centering
\caption{
Quantum number assignments, masses and strong partial decay widths in MeV of sextet $\Xi'_c$ (top) and anti-triplet $\Xi_c$ baryons (bottom) for the states reported in Table~\ref{XiQ}. Partial widths denoted by $-$ and $0$ are forbidden by 
phase space and seletion rules, respectively.}
\vspace{10pt}
\label{Xictable}
\begin{tabular}{cccccccccccccc}
\hline\hline
\noalign{\smallskip}
State & $M_{\rm th}$ & Baryon & $M_{\rm exp}$ & $^2\Sigma_c \bar{K}$ 
& $^2\Xi'_c \pi$ & $^4\Sigma_c \bar{K}$ & $^4\Xi'_c \pi$ & $\Lambda_c \bar{K}$ 
& $\Xi_c \pi$ & $\Xi_c \eta$ & $\Gamma_{\rm tot}$ & $\Gamma_{\rm exp}$ & Ref. \\
\noalign{\smallskip}
\hline
\noalign{\smallskip}
$^2\Xi'_c$ & $2570 \pm 2$ & $\Xi^{'+}_c$ & $2578.2 \pm 0.5$
& $-$  & $-$  & $-$  & $-$ & $-$  & $-$  & $-$  & $-$  & $-$  & \cite{PDG} \\
& & $\Xi^{'0}_c$ & $2578.7 \pm 0.5$
& $-$  & $-$  & $-$  & $-$ & $-$  & $-$  & $-$  & $-$  & $-$ & \cite{PDG} \\
$^{4}\Xi'_c$ & $2635 \pm 2$ & $\Xi_c(2645)^+$ & $2645.10 \pm 0.30$ & $-$  & $-$  
& $-$  & $-$  & $-$  & 0.39 & $-$  & 0.39 & $2.14 \pm 0.19$ & \cite{PDG} \\
&& $\Xi_c(2645)^0$ & $2646.16 \pm 0.25$ & $-$  & $-$  
& $-$  & $-$  & $-$  & 0.39 & $-$  & 0.39 & $2.35 \pm 0.22$ & \cite{PDG} \\
$^{2}\lambda(\Xi'_c)_{1/2^-}$ & $2905 \pm  2$ &&
& $-$  & 0.09 & $-$  & 0.18 & 0.02 & 0.67 & $-$  & 0.96 && \\
$^{4}\lambda(\Xi'_c)_{1/2^-}$ & $2934 \pm 3$ & $\Xi_c(2923)^0$ & $2923.04 \pm 0.35$ 
& $-$  & 0.11 & $-$  & 0.16 & 0.30 & 1.81 & $-$  & 2.38 & $7.1 \pm 2.0$ 
& \cite{Aaij:2020yyt} \\
$^{2}\lambda(\Xi'_c)_{3/2^-}$ & $2941 \pm 2$ & $\Xi_c(2930)^+$ & $2942.3 \pm 4.6$  
& $-$  & 1.76 & $-$  & 0.18 & 1.77 & 2.96 & $-$  & 6.68 & $14.8 \pm 9.1$ & \cite{PDG} \\
&& $\Xi_c(2939)^0$ & $2938.55 \pm 0.30$ & $-$  & 1.76 & $-$  & 0.18 & 1.77 & 2.96 & $-$  & 6.68 & $10.2 \pm 1.4$ & \cite{Aaij:2020yyt} \\ 
$^{4}\lambda(\Xi'_c)_{3/2^-}$ & $2970 \pm 2$ & $\Xi_c(2965)^0$ & $2964.88 \pm 0.33$
& 0.01 & 0.12 & $-$  & 1.00 & 0.44 & 0.67 & $-$  & 2.23 & $14.1 \pm 1.6$ 
& \cite{Aaij:2020yyt} \\
$^{4}\lambda(\Xi'_c)_{5/2^-}$ & $3030 \pm 2$ &&
& 0.66 & 1.07 & 0.05 & 2.30 & 3.39 & 4.72 & 0.02 & 12.20 && \\
$^{2}\rho(\Xi'_c)_{1/2^-}$ & $3060 \pm 2$ & $\Xi_c(3055)^+$ & $3055.9 \pm 0.4$ 
& 0.01 & 1.09 & 1.04 & 4.35 & $0$  & $0$  & $0$  & 6.49 & $7.8 \pm 1.9$ & \cite{PDG} \\
$^{2}\rho(\Xi'_c)_{3/2^-}$ & $3096 \pm 2$ & $\Xi_c(3080)^+$ & $3077.2 \pm 0.4$ 
& 5.01 & 3.66 & 1.89 & 3.58 & $0$  & $0$  & $0$  & 14.14 & $3.6 \pm 1.1$ 
& \cite{PDG} \\
&& $\Xi_c(3080)^0$ & $3079.9 \pm 1.4$ & 5.01 & 3.66 & 1.89 & 3.58 & $0$  & $0$  & $0$  & 14.14 & $5.6 \pm 2.2$ & \cite{PDG} \\
\noalign{\smallskip}
\hline
\noalign{\smallskip}
$^{2}\Xi_c$ & $2461 \pm 1$ & $\Xi_c^+$ & $2467.71 \pm 0.23$ 
& $-$  & $-$  & $-$  & $-$  &  $-$  & $-$  &  $-$ &  $-$ & $-$ & \cite{PDG} \\
&& $\Xi_c^0$ & $2470.44 \pm 0.28$ & $-$  & $-$  & $-$  & $-$  &  $-$  & $-$  &  $-$ &  $-$ &  $-$ & \cite{PDG} \\
$^{2}\lambda(\Xi_c)_{1/2^-}$ & $2797 \pm 1$ & $\Xi_c(2790)^+$ & $2791.9 \pm 0.5$ 
& $-$  & 0.02 & $-$  & 0.00 & $0$  & $0$  & $0$  & 0.02 & $8.9 \pm 1.0$ 
& \cite{PDG} \\
&& $\Xi_c(2790)^0$ & $2793.9 \pm 0.5$ & $-$  & 0.02 & $-$  & 0.00 & $0$  & $0$  & $0$  & 0.02 & $10.0 \pm 1.1$ & \cite{PDG} \\
$^{2}\lambda(\Xi_c)_{3/2^-}$ & $2832 \pm 1$ & $\Xi_c(2815)^+$ & $2816.51 \pm 0.25$ 
& $-$  & 0.23 & $-$  & 0.06 & $0$  & $0$  & $0$  & 0.29 & $2.43 \pm 0.26$ 
& \cite{PDG} \\
&& $\Xi_c(2815)^0$ & $2819.79 \pm 0.30$ & $-$  & 0.23 & $-$  & 0.06 & $0$  & $0$  & $0$  & 0.29 & $2.54 \pm 0.25$ & \cite{PDG} \\
$^{2}\rho(\Xi_c)_{1/2^-}$ & $2951 \pm 1$ &&
& 0.79 & 0.33 & $-$  & 0.43 & 0.26 & 1.08 & $-$  & 2.90 && \\
$^{4}\rho(\Xi_c)_{1/2^-}$ & $2980 \pm 2$ &&
& 0.69 & 0.28 & $-$  & 0.33 & 0.98 & 2.68 & $-$  & 4.96 && \\
$^{2}\rho(\Xi_c)_{3/2^-}$ & $2987 \pm 1$ &&  
& 0.44 & 2.75 & $-$  & 0.40 & 2.48 & 3.63 & $-$  & 9.69 && \\
$^{4}\rho(\Xi_c)_{3/2^-}$ & $3016 \pm 2$ &&  
& 0.08 & 0.17 & 0.99 & 1.99 & 0.56 & 0.78 & $-$  & 4.57 && \\
$^{4}\rho(\Xi_c)_{5/2^-}$ & $3076 \pm 2$ &&  
& 1.59 & 1.37 & 1.40 & 3.41 & 3.70 & 5.09 & 0.07 & 16.63 && \\
\noalign{\smallskip}
\hline\hline
\end{tabular}
\end{table*}

In the EEM the corresponding operator is given by \cite{KI,BIL4,BIL3}
\ba
{\cal H}_s &=& \frac{1}{(2\pi)^{3/2} (2k_0)^{1/2}}
\sum_{j=1}^{3} X^M_{j} \left[
2g \, (\vec{s}_j \cdot \vec{k}) \mbox{e}^{-i \vec{k} \cdot \vec{r}_j} \right.
\nonumber\\
&& \qquad \qquad \left.
+ h \, \vec{s}_j \cdot
(\vec{p}_j \, \mbox{e}^{-i \vec{k} \cdot \vec{r}_j} +
\mbox{e}^{-i \vec{k} \cdot \vec{r}_j} \, \vec{p}_j) \right] ~, 
\label{hs}
\ea
where $\vec{r}_j$, $\vec{p}_j$ and $\vec{s}_j$ are the coordinate,
momentum and spin of the $j$-th constituent, respectively;
$k_0$ is the meson energy and $\vec{k}=k \hat z$ denotes the momentum
carried by the outgoing meson. The flavor operator $X^M_j$ corresponds to the emission of an elementary meson by the $j$-th constituent:
$q_j \rightarrow q_j^{\prime} + M$. 
The coefficients $g$ and $h$ are fitted to two strong decays of $\Omega_c$ baryons
\ba 
\Gamma(\Omega_c(3050) \rightarrow \Xi_c + \bar{K}) &=& 0.8 \pm 0.2 ~,
\nonumber\\  
\Gamma(\Omega_c(3066) \rightarrow \Xi_c + \bar{K}) &=& 3.5 \pm 0.4 ~,
\ea   
to obtain $g=1.821$ GeV$^{-1}$ and $h=-0.356 $ GeV$^{-1}$ in qualitative agreement with values used in the light baryon sector \cite{BIL4,BIL3}. 

A comparison between the experimental data and the predicted mass spectra and strong  partial decay  widths calculated within  EEM are shown in Tables~\ref{Xictable} and \ref{Xibtable}. The zeros in the tables are consequences of the spin-flavor symmetry of the configurations $^2\rho(\Xi'_Q)_J$ and $^2\lambda(\Xi_Q)_J$. The calculated values are  of the order of 0-15 MeV in qualitative agreement with the experimental data.  

\begin{table*}
\centering
\caption{As in Table~\ref{Xictable}, but for sextet $\Xi'_b$ (top) and anti-triplet $\Xi_b$ baryons (bottom).}
\vspace{10pt}
\label{Xibtable}
\begin{tabular}{ccccccccccccccc}
\hline\hline
\noalign{\smallskip}
State & Assignment & $M_{\rm th}$  & $M_{\rm exp}$ & $^2\Sigma_b \bar{K}$ 
& $^2\Xi'_b \pi$ & $^4\Sigma_b \bar{K}$ & $^4\Xi'_b \pi$ & $\Lambda_b \bar{K}$ 
& $\Xi_b \pi$ & $\Xi_b \eta$ & $\Gamma_{\rm tot}$ & $\Gamma_{\rm exp}$ & Ref. \\
\noalign{\smallskip}
\hline
\noalign{\smallskip}
$^{2}\Xi'_b$ &     $\Xi'_b(5935)^-$   & $5926 \pm 2$ &   $5935.02 \pm 0.05$ 
& $-$  & $-$  & $-$  & $-$  & $-$  & $-$  & $-$  & $-$ & $< 0.08$ & \cite{PDG} \\

$^{4}\Xi'_b$ & $\Xi_b(5945)^0$  &  $5946 \pm 6$ &   $5952.3 \pm 0.6$ & $-$  & $-$  
& $-$  & $-$  & $-$  & 0.16 & $-$  & 0.16 & $0.90 \pm 0.18$ & \cite{PDG} \\ 

& $\Xi_b(5955)^-$  &  &   $5955.33 \pm 0.13$ & $-$  & $-$  
& $-$  & $-$  & $-$  & 0.16 & $-$  & 0.16 & $1.65 \pm 0.33$ & \cite{PDG} \\

$^{2}\lambda(\Xi'_b)_{1/2^-}$ &   & $6190 \pm  2$ &
& $-$  & 0.00 & $-$  & 0.15 & 0.03 & 0.66 & $-$  & 0.85 && \\

$^{4}\lambda(\Xi'_b)_{1/2^-}$ &   &  $6203 \pm 7$ &
& $-$  & 0.00 & $-$  & 0.11 & 0.00 & 1.59 & $-$  & 1.70 && \\  

$^{2}\lambda(\Xi'_b)_{3/2^-}$ & &  $6198 \pm 2$ &
& $-$  & 0.62 & $-$  & 0.10 & 1.07 & 2.80 & $-$  & 4.59 && \\

$^{4}\lambda(\Xi'_b)_{3/2^-}$ & &  $6210 \pm 6$ &
& $-$  & 0.04 & $-$  & 0.42 & 0.27 & 0.60 & $-$  & 1.33 && \\

$^{4}\lambda(\Xi'_b)_{5/2^-}$ &  $\Xi_b(6227)^0$ &  $6223 \pm 6$ &   $6226.8^{+1.5}_{-1.6}$ & $-$  & 0.30 & $-$  & 0.72 & 2.03 & 3.90 & $-$  
& 6.95 & $18.6^{+5.2}_{-4.3}$ & \cite{xibzero} \\

& $\Xi_b(6227)^-$ &&  $6227.9 \pm 0.9$ & $-$  & 0.30 & $-$  & 0.72 & 2.03 & 3.90 & $-$  
& 6.95 & $19.9 \pm 2.6$ & \cite{Aaij:2018yqz} \\

$^{2}\rho(\Xi'_b)_{1/2^-}$ &  & $6356 \pm 2$ &
& 0.55 & 0.66 & 0.44 & 3.92 & $0$  & $0$  & $0$  & 5.57 && \\

$^{2}\rho(\Xi'_b)_{3/2^-}$ & &  $6364 \pm 2$ &
& 1.06 & 2.32 & 1.27 & 2.64 & $0$  & $0$  & $0$  & 7.29 && \\
\noalign{\smallskip}
\hline
\noalign{\smallskip}
$^2\Xi_b$ & $\Xi_b^0$ &  $5784 \pm 2$  & $5791.9 \pm 0.5$ 
& $-$  & $-$  & $-$  & $-$ & $-$  & $-$  & $-$  & $-$  & $-$ & \cite{PDG} \\

& $\Xi_b^-$   &  &  $5797.0 \pm 0.6$ 
& $-$  & $-$  & $-$  & $-$ & $-$  & $-$  & $-$  & $-$  & $-$ & \cite{PDG} \\

$^{2}\lambda(\Xi_b)_{1/2^-}$ &   &  $6048 \pm 2$ &
& $-$  & $-$  & $-$  & $-$ & $0$  & $0$  & $0$  & $-$  & \\

$^{2}\lambda(\Xi_b)_{3/2^-}$ && $6056 \pm 2$ &
& $-$  & $-$  & $-$  & $-$ & $0$  & $0$  & $0$  & $-$  & \\

$^{2}\rho(\Xi_b)_{1/2^-}$ & &  $6214 \pm 2$ &
& $-$  & 0.01 & $-$  & 0.20 & 0.01 & 0.65 & $-$  & 0.87 & \\

$^{4}\rho(\Xi_b)_{1/2^-}$ && $6226 \pm 6$ &
& $-$  & 0.01 & $-$  & 0.13 & 0.09 & 1.48 & $-$  & 1.71 & \\

$^{2}\rho(\Xi_b)_{3/2^-}$ && $6222 \pm 2$ &
& $-$  & 0.71 & $-$  & 0.12 & 1.19 & 2.30 & $-$  & 4.32 & \\

$^{4}\rho(\Xi_b)_{3/2^-}$ && $6234 \pm 6$ &
& $-$  & 0.04 & $-$  & 0.50 & 0.28 & 0.49 & $-$  & 1.31 & \\

$^{4}\rho(\Xi_b)_{5/2^-}$ && $6247 \pm 6$ &
& $-$  & 0.31 & $-$  & 0.81 & 1.93 & 3.10 & $-$  & 6.16 & \\
\noalign{\smallskip}
\hline\hline
\end{tabular}
\end{table*}

In the $^{3}P_0$ model the transition operator is given by \cite{Micu:1968mk,LeYaouanc:1972vsx,LeYaouanc:1988fx,Roberts:1997kq,Chen:2007xf,Bijker}
\begin{eqnarray}
T^{\dagger} &=& -3 \gamma_0 \, \int d \vec{p}_4 \, d \vec{p}_5 \, 
\delta(\vec{p}_4 + \vec{p}_5) \, C_{45} \, F_{45} 
\nonumber\\
&& \hspace{0.5cm}  \left[ \chi_{45} \, \times \, {\cal Y}_{1}(\vec{p}_4 - \vec{p}_5) \right]^{(0)}_0 \, 
b_4^{\dagger}(\vec{p}_4) \, d_5^{\dagger}(\vec{p}_5)    \mbox{ }.
\label{3p0}
\end{eqnarray}
Here, $\gamma_0$ is the pair-creation strength, and $b_4^{\dagger}(\vec{p}_4)$ and $d_5^{\dagger}(\vec{p}_5)$ are the creation operators for a quark and an antiquark with momenta $\vec{p}_4$ and $\vec{p}_5$, respectively.
The $q \bar q$ pair is characterized by a color-singlet wave function $C_{45}$, a flavor-singlet wave function $F_{45}$, a spin-triplet wave function $\chi_{45}$ with spin $S=1$ and a solid spherical harmonic ${\cal Y}_{1}(\vec{p}_4 - \vec{p}_5)$, since the quark and antiquark are in a relative $P$-wave. 
In Ref.~\cite{Santopinto:2018ljf}, the pair-creation strength, $\gamma_0$, was fitted to the charge channel $\Omega_c(3066) \rightarrow \Xi^+_c K^-$. Here, we fit $\gamma_0$ to the isospin channel $\Xi_c \bar{K}$ to obtain $\gamma_0=9.22/\sqrt{2}=6.52$. 

\begin{table}[t]
\centering
\caption{Comparison between the strong total decay widths for charmed baryons $\Xi_c$ and $\Xi_c^{'}$  calculated within EEM and $^3P_0$ model with  the chiral quark model \cite{Wang:2017kfr} and the available experimental data. All values are expressed in MeV.}
\label{dwc}
\vspace{10pt}
\begin{tabular}{crrccc}
\hline\hline
\noalign{\smallskip}
& \multicolumn{2}{c}{Present} & $\chi$QM & Exp & Baryon \\
State & EEM & $^3P_0$ & \cite{Wang:2017kfr}& \cite{PDG,Aaij:2020yyt} & \\
\noalign{\smallskip}
\hline 
\noalign{\smallskip}
$^4\Xi'_{c}$                               &  0.39 &  0.02 &           & $2.14 \pm 0.19$ 
& $\Xi_c(2645)^+$ \\                   & 0.39 &  0.02 &           & $2.35 \pm 0.22$ & $\Xi_c(2645)^0$ \\
$^2\lambda(\Xi'_{c})_{{1/2}^-}$ &  0.96 &  0.79 & 21.67 && \\
$^4\lambda(\Xi'_{c})_{{1/2}^-}$ &  2.38 &  0.53 & 37.05 & $ 7.1 \pm 2.0$ 
& $\Xi_c(2923)^0$ \\
$^2\lambda(\Xi'_{c})_{{3/2}^-}$ &  6.68 &  3.08 & 20.89 & $14.8 \pm 9.1$ 
& $\Xi_c(2930)^+$ \\                  & 6.68 &    3.08 && $10.2 \pm 1.4$ & $\Xi_c(2939)^0$ \\
$^4\lambda(\Xi'_{c})_{{3/2}^-}$ &  2.23 &  2.04 & 12.33 & $14.1 \pm 1.6$ 
& $\Xi_c(2965)^0$ \\
$^4\lambda(\Xi'_{c})_{{5/2}^-}$ & 12.20 &  5.43 & 20.20 &&  \\
$^2\rho(\Xi'_{c})_{{1/2}^-}$      &  6.49  &   6.26&       & $ 7.8 \pm 1.9$ 
& $\Xi_c(3055)^+$ \\
$^2\rho(\Xi'_{c})_{{3/2}^-}$    & 14.14 &  3.70 &       & $ 3.6 \pm 1.1$ 
& $\Xi_c(3080)^+$ \\               & 14.14  & 3.70 && $5.6 \pm 2.2$ & $\Xi_c(3080)^0$ \\
\noalign{\smallskip}
\hline
\noalign{\smallskip}
$^2\lambda(\Xi_{c})_{{1/2}^-}$ &  0.02 &  0.41 & 3.61 & $8.9  \pm 1.0$ 
& $\Xi_c(2790)^+$ \\                 & 0.02 & 0.41 && $10.0 \pm 1.1$ & $\Xi_c(2790)^0$ \\
$^2\lambda(\Xi_{c})_{{3/2}^-}$ &  0.29 &  0.54 & 2.11 & $2.43 \pm 0.26$ 
& $\Xi_c(2815)^+$ \\                 & 0.29 & 0.54 && $2.54 \pm 0.25$ & $\Xi_c(2815)^0$\\
$^2\rho(\Xi_{c})_{{1/2}^-}$    &  2.90 & 0.70 &&& \\
$^4\rho(\Xi_{c})_{{1/2}^-}$    &  4.96 &  0.45 &&& \\
$^2\rho(\Xi_{c})_{{3/2}^-}$    &  9.69 &  3.76 &&& \\
$^4\rho(\Xi_{c})_{{3/2}^-}$    &  4.57 & 2.51 &&& \\
$^4\rho(\Xi_{c})_{{5/2}^-}$    & 16.63 & 4.55&&& \\
\noalign{\smallskip}
\hline\hline 
\end{tabular}
\end{table}

In Tables~\ref{Xictable} and \ref{Xibtable} we show the two-body decay widths calculated in the elementary emission model. A comparison of the mass spectrum and the strong decay widths with experimental results suggests that the $\Xi_{c}(2923)$, $\Xi_{c}(2939)$, and $\Xi_{c}(2965)$ baryons are negative parity $P$-wave states of $\Xi'_c$ and/or $\Xi_c$ states.
In fact, we can summarize the emerging possible identification of the three new $\Xi_{c}$ states observed by LHCb \cite{Aaij:2020yyt} and the $\Xi_b(6227)$ reported in \cite{Aaij:2018yqz} as negative parity $P$-wave $\Xi'_Q$ $\lambda$-excitation states of the flavor sextet $\bf 6$
\begin{eqnarray}
^{4}\lambda(\Xi'_c)_{1/2^-} &\rightarrow& \Xi_{c}(2923) 
\nonumber\\
^{2}\lambda(\Xi'_c)_{3/2^-} &\rightarrow& \Xi_{c}(2939)
\nonumber\\
^{4}\lambda(\Xi'_c)_{3/2^-} &\rightarrow& \Xi_{c}(2965) 
\nonumber\\
^{4}\lambda(\Xi'_b)_{5/2^-} &\rightarrow& \Xi_{b}(6227) 
\end{eqnarray}
and/or as negative parity $P$-wave $\Xi_Q$ $\rho$-excitation states of the  
flavor anti-triplet $\bf \bar{3}$
\begin{eqnarray}
^{2}\rho(\Xi_c)_{1/2^-} &\rightarrow& \Xi_{c}(2939)
\nonumber\\
^{4}\rho(\Xi_c)_{1/2^-} &\rightarrow& \Xi_{c}(2965) 
\nonumber\\
^{2}\rho(\Xi_b)_{3/2^-} &\rightarrow& \Xi_{b}(6227) 
\end{eqnarray}
Each of those states satisfies the equal spacing rules with the $\Omega_c$ or $\Omega_b$ states: the $P$-wave $\Xi_Q$ $\lambda$-excitation states since they belong to the same ${\bf 6}$-plet, while the $P$-wave $\Xi_Q$ $\rho$-excitation states that belong to a $\bf{ \bar 3}$-plet due to accidental degeneration in the spectrum. 
We can not exclude a priori that the states seen by LHCb corresponds to all of them, also because each of those states can decay into $\Lambda_c \bar{K}$ channel.
LHCb, BELLE and BaBar can do new analysis to test if the $P$-wave $\Xi_Q^{(')}$ are 14 states or not.
The future amplitude analysis and the subsequently measurement of their $J^P$ quantum numbers will be crucial in order to disentangle the correct identification of those states. 
The fact that in our model $\Xi_c$(3055) and $\Xi_c$(3080) resonances are assigned as $\rho$-mode 
excitations of the sextet configuration, $^2\rho(\Xi'_c)_{J^P}$ with $J^P=1/2^-$ and $3/2^-$,  indicates a preference of a three-quark over a quark-diquark description. However, this assignment is mainly based on the predicted mass spectrum and we cannot exclude that all the five states, 
$\Xi_{c}(2923),\Xi_{c}(2939),\Xi_{c}(2965),  \Xi_{c}(3055)$ and $\Xi_{c}(3080)$ are instead 
$\lambda$-mode  excitations:
\begin{eqnarray}
^{2}\lambda(\Xi_c^{'})_{1/2^-} &\rightarrow& \Xi_{c}(2923)
\nonumber\\
^{4}\lambda(\Xi_c^{'})_{1/2^-} &\rightarrow& \Xi_{c}(2939) 
\nonumber\\
^{2}\lambda(\Xi_c^{'})_{3/2^-}  &\rightarrow&  \Xi_{c}(2965) 
\nonumber\\
^{4}\lambda(\Xi_c^{'})_{3/2^-}  &\rightarrow&  \Xi_{c}(3055) 
\nonumber\\
^{4}\lambda(\Xi_c^{'})_{5/2^-}  &\rightarrow&  \Xi_{c}(3080) 
\end{eqnarray}


In Tables~\ref{dwc} and \ref{dwb} we present the total strong decay widths 
calculated in the elementary emission model and the $^3P_0$ model for the harmonic oscillator quark model, and show a comparison with the chiral quark model \cite{Wang:2017kfr} and the available experimental data. 
In general the values obtained in \cite{Wang:2017kfr} are too large in comparison with the experimental widths.  
The predicted widths for $\Xi_b(6327)^{0}$ and $\Xi_b(6333)^{0}$ are 0.19 MeV and 0.10 MeV  respectively,  and they are in 
 good agreement with the experimental width measured by the LHCb collaboration \cite{LHCb:2021ssn}: $ \Gamma\left(\Xi_{b}(6327)^{0}\right) = 0.93_{-0.60}^{+0.74} $  MeV and $\Gamma\left(\Xi_{b}(6333)^{0}\right) = 0.25_{-0.25}^{+0.58} $ MeV, See Table \ref{dwb}.  

\begin{table}[h]
\centering
\caption{As in Tab. \ref{dwc} but for $\Xi_b$ and $\Xi_b^{'}$.}
\label{dwb}
\vspace{10pt}
\begin{tabular}{crrccc}
\hline\hline
\noalign{\smallskip}
& \multicolumn{2}{c}{Present} & $\chi$QM & Exp & Baryon \\
State & EEM & $^3P_0$ & \cite{Wang:2017kfr} & \cite{PDG,Aaij:2018yqz,xibzero,LHCb:2021ssn} \\
\noalign{\smallskip}
\hline 
\noalign{\smallskip}
$^2\Xi'_{b}$                    & $-$  & $-$   &  0.08 & $<0.08$ \\
$^4\Xi'_{b}$                    & 0.16 &  0.02 &  0.98 & $0.90 \pm 0.18$ 
& $\Xi_b(5945)^0$ \\ & 0.16 & 0.02 && $1.65 \pm 0.33$ & $\Xi_b(5955)^-$ \\
$^2\lambda(\Xi'_{b})_{{1/2}^-}$ & 0.85 &   0.86 & 27.05 && \\
$^4\lambda(\Xi'_{b})_{{1/2}^-}$ & 1.70 &   0.65 & 32.24 && \\
$^2\lambda(\Xi'_{b})_{{3/2}^-}$ & 4.59 &   2.92 & 24.15 && \\
$^4\lambda(\Xi'_{b})_{{3/2}^-}$ & 1.33 &  1.83& 15.83 && \\
$^4\lambda(\Xi'_{b})_{{5/2}^-}$ & 6.95 &  3.36 & 24.39 & $18.6^{+5.2}_{-4.3}$ 
& $\Xi_b(6227)^0$ \\ 
& 6.95 &  3.36 && $19.9 \pm 2.6$ & $\Xi_b(6227)^-$ \\
$^2\rho(\Xi'_{b})_{{1/2}^-}$    & 5.57 & 5.88 &&& \\
$^2\rho(\Xi'_{b})_{{3/2}^-}$    & 7.29 &  3.08 &&& \\
\noalign{\smallskip}
\hline
\noalign{\smallskip}
$^2\lambda(\Xi_{b})_{{1/2}^-}$ & $-$  & $-$    & 2.88 && \\
$^2\lambda(\Xi_{b})_{{3/2}^-}$ & $-$  & $-$    & 2.95 && \\
$^2\rho(\Xi_{b})_{{1/2}^-}$    & 0.87 &  0.55 &&& \\
$^4\rho(\Xi_{b})_{{1/2}^-}$    & 1.71 &  0.36 &&& \\
$^2\rho(\Xi_{b})_{{3/2}^-}$    & 4.32 & 1.90 &&& \\
$^4\rho(\Xi_{b})_{{3/2}^-}$    & 1.31 & 1.90 &&& \\
$^4\rho(\Xi_{b})_{{5/2}^-}$    & 6.16 & 2.16 &&& \\
$D$-wave                              &&&&&\\
$^2 \lambda(\Xi_{b})_{{3/2}^+}$    &  & 0.19 &&$0.93^{+0.74}_{-60}$&$\Xi_b(6327)^0$ \\
$^2\lambda(\Xi_{b})_{{5/2}^+}$    &  & 0.10 &&$0.25^{+0.58}_{-25}$&$\Xi_b(6333)^0$ \\

\noalign{\smallskip}
\hline\hline 
\end{tabular}
\end{table}

\subsection{Electromagnetic couplings}

In constituent models, electromagnetic couplings arise from the 
coupling of the (point-like) constituent parts to the electromagnetic 
field \cite{copley1,copley2,BIL1,BIL2}. We discuss here the 
case of the emission of a lefthanded photon 
\ba
B \rightarrow B' + \gamma ~, 
\ea
for which the nonrelativistic part of the transverse 
electromagnetic coupling is given by
\ba
{\cal H}_{em} &=& 2 \sqrt{\frac{\pi}{k_0}} \sum_{j=1}^{3} \mu_j e_j \, 
\left[ k s_{j,-} \, \mbox{e}^{-i \vec{k} \cdot \vec{r}_j} \right.
\nonumber\\
&& \qquad \left. + \frac{1}{2g_j} (p_{j,-} \, \mbox{e}^{-i \vec{k} \cdot \vec{r}_j} +
\mbox{e}^{-i \vec{k} \cdot \vec{r}_j} \, p_{j,-}) \right] ~, 
\ea
where $\vec{r}_j$, $\vec{p}_j$ and $\vec{s}_j$ are the coordinate,
momentum and spin of the $j$-th constituent, respectively;
$k_0$ is the photon energy, and $\vec{k}=k \hat z$ 
denotes the momentum carried by the outgoing photon.  
The photon is emitted by the $j$-th constituent: 
$q_j \rightarrow q_j^{\prime} + \gamma$.

\begin{table*}[h]
\centering
\caption{Radiative decay widths of ground states $\Xi'_Q$ and $\Xi_Q$ baryons in KeV.}
\vspace{10pt}
\label{radgs}
\begin{tabular}{lcccccccccccc}
\hline\hline
\noalign{\smallskip}
Decay & $q$ & Present & LCQSR & BM & VDM & $\chi$QM & NRQM & HB$\chi$PT & RQM & hCQM \\
&&& \cite{Aliev2009}-\cite{Aliev2016} & \cite{Bernotas2013} 
& \cite{Aliev2012} & \cite{Wang:2017kfr} & \cite{Majethiya2009} 
& \cite{JuanWang2019} & \cite{Ivanov1999} & \cite{Shah2018} \\
\noalign{\smallskip}
\hline
\noalign{\smallskip}
$^2\Xi'_{c} \rightarrow {}^2\Xi_{c} +\gamma$ 
& $+$ & 14.2 & $8.5 \pm 2.5$ & 10.2 & & 42.3 & & $5.43 \pm 0.33$ & $12.7 \pm 1.5$ & \\
& $0$ & 0.3 & $0.27 \pm 0.06$ & 0.0015 & & 0.00 & & 0.46 & $0.17 \pm 0.002$ \\
$^4\Xi'_{c} \rightarrow {}^2\Xi_{c} +\gamma$ 
& $+$ & 51.1 & $52 \pm 32$ & 44.3 & 152.4 & 139 & 63.32 & $21.6 \pm 1$ & $54 \pm 3$ & 17.48 \\
& $0$ & 1.1 & $0.66 \pm 0.41$ & 0.908 & 1.318 & 0.00 & 0.30 & 1.84 & $0.68 \pm 0.04$ & 0.45 \\
$^4\Xi'_{c} \rightarrow {}^2\Xi'_{c} +\gamma$ 
& $+$ & 0.0 & 0.274 & 0.011 & 0.485 & 0.004 & & $0.07 \pm 0.03$ & & \\
& $0$ & 0.9 & 2.142 & 1.03  & 1.317 & 3.03  & & $0.42 \pm 0.16$ & & \\
\noalign{\smallskip}
\hline
\noalign{\smallskip}
$^2\Xi'_{b} \rightarrow {}^2\Xi_{b} +\gamma$ 
& $0$ & 29.5 & $47.0 \pm 21.0$ & 14.7 & & 84.6 & & $13.0 \pm 0.8$ \\
& $-$ & 0.6 & $3.3 \pm 1.3$ & 0.118 & & 0.00 & & 1.0 \\
$^4\Xi'_{b} \rightarrow {}^2\Xi_{b} +\gamma$ 
& $0$ & 42.4 & $ 135 \pm 85$   & 24.7 & 270.8 & 104 & 18.79 & $17.2 \pm 0.1$ \\
& $-$ &  0.9 & $1.50 \pm 0.95$ & 0.278 & 2.246 & 0.00 & 0.09 & 1.4  \\
$^4\Xi'_{b}\rightarrow {}^2\Xi'_{b} +\gamma$ 
& $0$ & 0.0 & 0.131 & 0.004 & 0.281 & 5.19 & & $(1.5 \pm 0.5) \times 10^{-3}$ \\
& $-$ & 0.0 & 0.303 & 0.005 & 0.702 & 15.0 & & $(8.2 \pm 4) \times 10^{-3}$ \\
\noalign{\smallskip}
\hline\hline
\end{tabular}
\end{table*}

\begin{table}[t]
\centering
\caption{Radiative decay widths of excited $P$-wave $\Xi'_c$ and $\Xi_c$ baryons in KeV.}
\vspace{10pt}
\label{radexcc}
\begin{tabular}{crrrrrrl}
\hline\hline
\noalign{\smallskip}
& \multicolumn{2}{c}{$^2\Xi'_{c} +\gamma$} 
& \multicolumn{2}{c}{$^4\Xi'_{c} +\gamma$} 
& \multicolumn{2}{c}{$^2\Xi_{c} +\gamma$} & \\ 
& $+$ & $0$ & $+$ & $0$ & $+$ & $0$ & \\ 
\noalign{\smallskip}
\hline
\noalign{\smallskip}
$^2\lambda(\Xi'_{c})_{1/2^-}$ & 11.9 & 451.0 & 1.0 &   0.4 &  53.3 & 1.1 & \\
                              &  0.0 & 472.0 & 1.6 &   1.0 &  46.4 & 0.0 & \cite{Wang:2017kfr} \\
$^2\lambda(\Xi'_{c})_{3/2^-}$ & 12.0 & 850.3 & 1.7 &   0.6 &  60.8 & 1.3 & \\
                              & 12.1 & 302.0 & 1.6 &   1.0 &  46.1 & 0.0 & \cite{Wang:2017kfr} \\
$^4\lambda(\Xi'_{c})_{1/2^-}$ &  3.2 &   0.9 & 2.5 &  54.2 &  29.7 & 0.6 & \\
                              &  0.3 &   0.2 & 0.2 & 125.0 &  14.5 & 0.0 & \cite{Wang:2017kfr} \\
$^4\lambda(\Xi'_{c})_{3/2^-}$ & 12.1 &   2.7 & 4.2 & 296.4 &  91.7 & 1.9 & \\
                              &  2.1 &   1.2 & 1.6 & 187.0 &  54.6 & 0.0 & \cite{Wang:2017kfr} \\
$^4\lambda(\Xi'_{c})_{5/2^-}$ & 11.7 &   1.8 & 2.2 & 583.6 &  63.6 & 1.3 & \\
                              &  1.6 &   0.9 & 2.3 & 192.0 &  32.0 & 0.0 & \cite{Wang:2017kfr} \\
$^2\rho(\Xi'_{c})_{1/2^-}$    & 16.3 &  26.7 & 6.3 &  10.4 & 423.6 & 9.0 & \\
$^2\rho(\Xi'_{c})_{3/2^-}$    & 17.6 &  28.8 & 7.4 &  12.2 & 370.8 & 7.9 & \\
\noalign{\smallskip}
\hline
\noalign{\smallskip}
$^2\lambda(\Xi_{c})_{1/2^-}$ &   6.1 &  0.1 &   0.6 &  0.0 &  0.8 & 555.0 & \\
                             &   1.4 &  0.0 &   0.4 &  0.0 &  4.6 & 263.0 & \cite{Wang:2017kfr} \\
$^2\lambda(\Xi_{c})_{3/2^-}$ &  11.0 &  0.2 &   1.6 &  0.0 &  4.4 & 758.8 & \\
                             &   2.3 &  0.0 &   1.0 &  0.0 &  2.8 & 292.0 & \cite{Wang:2017kfr} \\
$^2\rho(\Xi_{c})_{1/2^-}$    & 251.5 &  5.3 &   3.6 &  0.1 & 16.2 &  26.5 & \\
                             & 128.0 &  0.0 &   0.2 &  0.0 &  1.4 &   5.6 & \cite{Wang:2017kfr} \\
$^2\rho(\Xi_{c})_{3/2^-}$    & 811.0 & 17.2 &   5.1 &  0.1 & 17.5 &  28.6 & \\
                             & 110.0 &  0.0 &   0.5 &  0.0 &  1.9 &   7.5 & \cite{Wang:2017kfr} \\
$^4\rho(\Xi_{c})_{1/2^-}$    &   7.8 &  0.2 &  23.0 &  0.5 &  8.6 &  14.1 & \\
                             &   0.4 &  0.0 &  43.4 &  0.0 &  0.7 &   3.0 & \cite{Wang:2017kfr} \\
$^4\rho(\Xi_{c})_{3/2^-}$    &  25.9 &  0.5 & 197.1 &  4.2 & 25.3 &  41.4 & \\
                             &   1.8 &  0.0 &  58.1 &  0.0 &  2.8 &  11.2 & \cite{Wang:2017kfr} \\
$^4\rho(\Xi_{c})_{5/2^-}$    &  20.1 &  0.4 & 561.2 & 11.9 & 16.3 &  26.7 & \\
\noalign{\smallskip}
\hline\hline
\end{tabular}
\end{table}

\begin{table}[t]
\centering
\caption{Radiative decay widths of excited $P$-wave $\Xi'_b$ and $\Xi_b$ 
baryons in KeV.}
\vspace{10pt}
\label{radexcb}
\begin{tabular}{crrrrrrl}
\hline\hline
\noalign{\smallskip}
& \multicolumn{2}{c}{$^2\Xi'_{b} +\gamma$} 
& \multicolumn{2}{c}{$^4\Xi'_{b} +\gamma$} 
& \multicolumn{2}{c}{$^2\Xi_{b} +\gamma$} & \\ 
& $0$ & $-$ & $0$ & $-$ & $0$ & $-$ & \\ 
\noalign{\smallskip}
\hline
\noalign{\smallskip}
$^2\lambda(\Xi'_{b})_{1/2^-}$ &  95.7 &  81.1 &   0.5 &   0.9 &  55.9 & 1.2 & \\
                              &  76.3 & 190.0 &   0.9 &   3.5 &  72.2 & 0.0 & \cite{Wang:2017kfr} \\
$^2\lambda(\Xi'_{b})_{3/2^-}$ & 192.7 & 201.7 &   0.6 &   1.0 &  57.9 & 1.2 & \\
                              &  43.9 &  92.3 &   0.9 &   3.6 &  72.8 & 0.0 & \cite{Wang:2017kfr} \\
$^4\lambda(\Xi'_{b})_{1/2^-}$ &   0.8 &   1.4 &  13.3 &   9.2 &  29.5 & 0.6 & \\
                              &   0.3 &   1.5 &  69.5 & 164.0 &  34.0 & 0.0 & \cite{Wang:2017kfr} \\
$^4\lambda(\Xi'_{b})_{3/2^-}$ &   2.4 &   4.3 &  70.2 &  61.9 &  84.8 & 1.8 & \\
                              &   0.7 &   2.0 &  47.5 & 104.0 &  94.0 & 0.0 & \cite{Wang:2017kfr} \\
$^4\lambda(\Xi'_{b})_{5/2^-}$ &   1.8 &   3.2 & 119.2 & 129.8 &  56.8 & 1.2 & \\
                              &   0.4 &   1.9 &  41.5 &  88.2 &  47.7 & 0.0 & \cite{Wang:2017kfr} \\
$^2\rho(\Xi'_{b})_{1/2^-}$    &  11.2 &  18.3 &   5.2 &   8.5 & 270.8 & 5.7 & \\
$^2\rho(\Xi'_{b})_{3/2^-}$    &  11.5 &  18.7 &   5.4 &   8.8 & 258.9 & 5.5 & \\
\noalign{\smallskip}
\hline
\noalign{\smallskip}
$^2\lambda(\Xi_{b})_{1/2^-}$ &   0.5 &  0.0 &   0.1 & 0.0 & 143.7 & 144.0 & \\
                             &   1.3 &  0.0 &   2.0 & 0.0 &  63.6 & 135.0 & \cite{Wang:2017kfr} \\
$^2\lambda(\Xi_{b})_{3/2^-}$ &   0.7 &  0.0 &   0.2 & 0.0 & 153.2 & 149.0 & \\
                             &   1.7 &  0.0 &   2.6 & 0.0 &  68.3 & 147.0 & \cite{Wang:2017kfr} \\
$^2\rho(\Xi_{b})_{1/2^-}$    & 207.5 &  4.4 &   2.1 & 0.0 &  11.0 &  18.0 & \\
                             &  94.3 &  0.0 &   0.6 & 0.0 &   1.9 &   7.2 & \cite{Wang:2017kfr} \\
$^2\rho(\Xi_{b})_{3/2^-}$    & 494.7 & 10.5 &   2.4 & 0.1 &  11.3 &  18.6 & \\
                             &  69.4 &  0.0 &   0.8 & 0.0 &   2.1 &   8.1 & \cite{Wang:2017kfr} \\
$^4\rho(\Xi_{b})_{1/2^-}$    &   3.1 &  0.1 &  24.1 & 0.5 &   5.7 &   9.4 & \\
                             &   0.2 &  0.0 &  80.0 & 0.0 &   0.9 &   3.6 & \cite{Wang:2017kfr} \\
$^4\rho(\Xi_{b})_{3/2^-}$    &   9.4 &  0.2 & 156.3 & 3.3 &  16.4 &  26.8 & \\
                             &   0.8 &  0.0 &  78.0 & 0.0 &   2.9 &  11.4 & \cite{Wang:2017kfr} \\
$^4\rho(\Xi_{b})_{5/2^-}$    &   6.9 &  0.1 & 315.6 & 6.7 &  10.8 &  17.7 & \\
\noalign{\smallskip}
\hline\hline
\end{tabular}
\end{table}

\begin{table*}[t]
\centering
\caption{Radiative decay widths of $\Xi_c(2790)$ and $\Xi_c(2815)$ baryons in KeV.}
\vspace{10pt}
\label{emxic}
\begin{tabular}{crrccc}
\hline\hline
\noalign{\smallskip}
Decay & Present & $\chi$QM & MB & LCQSR & Exp \\
&& \cite{Wang:2017kfr} & \cite{Gamermann} & \cite{Aliev2019} & \cite{Yelton} \\
\noalign{\smallskip}
\hline
\noalign{\smallskip}
$\Xi_{c}^+(2790) \rightarrow {}^2\Xi_{c}^+ + \gamma$ 
& $0.8$   & $4.6$   & $249.6 \pm 41.9$ & $265 \pm 106$ & $<350$ \\
$\Xi_{c}^0(2790) \rightarrow {}^2\Xi_{c}^0 + \gamma$ 
& $555.0$ & $263.0$ & $119.3 \pm 21.7$ & $2.7 \pm 0.8$ & $800 \pm 320$ \\
$\Xi_{c}^+(2815) \rightarrow {}^2\Xi_{c}^+ + \gamma$ 
& $4.4$   & $2.8$   &&& $<80$ \\
$\Xi_{c}^0(2815) \rightarrow {}^2\Xi_{c}^0 + \gamma$ 
& $758.8$ & $292.0$ &&& $320 \pm 45^{+45}_{-80}$ \\
\noalign{\smallskip}
\hline\hline
\end{tabular}
\end{table*}

Radiative decay widths of ground states $\Xi'_Q$ and $\Xi_Q$ baryons,
excited $P$-wave $\Xi'_c$ and $\Xi_c$, and  excited $P$-wave $\Xi'_b$ and $\Xi_b$ are reported respectively in Tables \ref{radgs}, \ref{radexcc}, and \ref{radexcb}.
Recently, the electromagnetic decays widths of the $\Xi_c(2790)$ and $\Xi_c(2815)$ 
baryons were measured by the Belle Collaboration \cite{Yelton}. The decays widths 
of the neutral states were found to be large and of the order of several hundreds 
of KeV, whereas for the charged states only an upper limit could be established. 
Inspection of Table~\ref{radexcc} shows that this behavior is in agreement 
only with an interpretation in terms of a $\lambda$-mode excitation of 
the flavor anti-triplet $^{2}\lambda(\Xi_c)_J$. A similar result was found in 
the $\chi$QM \cite{Wang:2017kfr}. The results are summarized in Table~\ref{emxic}. 
In addition, Table~\ref{radexcc} shows that there are several 
other decay widths which are expected to be large, like for example 
\ba
^{2}\lambda(\Xi_c^{\prime 0})_J &\rightarrow& {}^2\Xi_c^{\prime 0} + \gamma ~,
\nonumber\\
^{4}\lambda(\Xi_c^{\prime 0})_J &\rightarrow& {}^4\Xi_c^{\prime 0} + \gamma ~,
\nonumber\\
^{2}\rho(\Xi_c^{\prime +})_J &\rightarrow& {}^2\Xi_c^{+} + \gamma ~,
\nonumber\\
^{2}\rho(\Xi_c^{+})_J &\rightarrow& {}^2\Xi_c^{\prime +} + \gamma ~,
\nonumber\\
^{4}\rho(\Xi_c^{+})_J &\rightarrow& {}^4\Xi_c^{\prime +} + \gamma ~,
\ea
as well as the corresponding decay widths in Table~\ref{radexcb} for the beauty $\Xi'_b$ and $\Xi_b$ baryons.

\section{Summary and conclusions}

This work was motivated by recent LHCb measurements of $\Xi_c^0$ baryons 
and the observation of the equal spacing rules with $\Omega_c$ resonances. 
We presented a quark model calculation of ground state and excited heavy 
$\Xi_Q$ baryons with $Q=c$ and $b$ involving both sextet ($\Xi'_Q$) and 
anti-triplet ($\Xi_Q$) states. According to the quark model analysis there 
are 14 negative parity $P$-wave states divided evenly between the sextet 
(5 $\lambda$-mode and 2 $\rho$-mode) and the anti-triplet (2 $\lambda$-mode 
and 5 $\rho$-mode). 
 
The mass spectrum was obtained in a harmonic oscillator quark model with 
spin, spin-orbit, isospin and flavor dependent terms. The parameters were 
taken from a previous study of $\Omega_c$ and $\Omega_b$ baryons. 
The assignments of quantum numbers was based on systematics of the 
mass spectrum as well as on the strong and electromagnetic decay widths. Just 
as in the case of the masses, the parameters in the elementary emission model 
and the $^{3}P_0$ model for strong decays were taken from a study of $\Omega_c$ 
baryons. No attempt was made to fine-tune the parameter values to the 
spectroscopic properties of the $\Xi'_c$ and $\Xi'_c$ baryons. 

Overall, there is a reasonable agreement with the available experimental. 
In particular, it was found that the electromagnetic decays widths of the 
neutral and charged $\Xi_c(2790)$ and $\Xi_c(2815)$ baryons is compatible 
only with an assignment as a $\lambda$-mode excitation of the 
$^{2}\lambda(\Xi_c)_J$ configuration of the flavor anti-triplet.  
At the moment, not for all states such an unambigous assignment can be made. 
We have identified several large electromagnetic decay widths of the order 
of several hundreds of KeV which may help future experimental studies of 
charm and bottom baryons. The future measurement of the $J^P$ quantum numbers is crucial to determine  
the correct assignment of those states. Moreover, the identification of the 
negative parity $P$-wave states of the $\rho$-mode excitation is important to distinguish between an interpretation in terms of a three-quark or a quark-diquark structure, since in a quark-diquark model the $\rho$-mode excitations are frozen  
\cite{Santopinto:2018ljf}.
In the bottom sector our model is able to reproduce the masses and widths of the  two  new   $\Xi_b$  states by LHCb \cite{LHCb:2021ssn}, which have been identified as positive parity D-wave excitations of the $\Xi_b$ system with the following quantum number assignment  $J^{P}_{\Xi_b(6327)^{0}}=\frac{3}{2}^{+}$ and $J^{P}_{\Xi_b(6333)^{0}}=\frac{5}{2}^{+}$.
In the charm sector by providing results for the spectrum of both $\Xi_c$ and  $\Xi^{\prime}_{c}$ baryons and combining them with predictions for their $\Lambda_{c}^{+}K^{-}$ decays and their total decay withs, we can suggest assignments that can be tested by further experimental amplitude analysis.

\section*{Acknowledgments}
This work was supported by INFN and by grant IN101320 from DGAPA-UNAM.

\end{document}